\documentclass{PoS}
\usepackage{macros}
\usepackage[utf8]{inputenc} 

\newcommand{\circled}[1]{\raisebox{.5pt}{\textcircled{\raisebox{0pt} {{\scriptsize #1}}}}}

\title{CP violation and Kaon weak matrix elements from Lattice QCD}

\ShortTitle{CP violation and Kaon weak matrix elements from Lattice QCD}

\author{\speaker{Nicolas Garron}%
         \thanks{Supported by the Leverhulme Research grant  RPG-2014-118}\\
Centre for Mathematical Sciences.
School of Computing, Electronics and Mathematics,
Plymouth University, Plymouth, PL4 8AA, United Kingdom  \\
        E-mail: \email{nicolas.garron@plymouth.ac.uk}}

\abstract{
In this short review, I present the recent 
lattice computations of 
kaon weak matrix elements relevant to $K\to \pi\pi$
decays and neutral kaon mixing. These matrix elements 
are key to the theoretical determination of the CP violation parameters 
$\epsilon$ and $\epsilon'$. 
Impressive progress have been achieved recently,
in particular the first realistic computation of $\epsilon'/\epsilon$
with physical kinematics has been reported in~\cite{Bai:2015nea}.
The novelty is the $\Delta I=1/2$ channel,
whereas the $\Delta I=3/2$ contribution is now computed at several values 
of the lattice spacing and extrapolated to the continuum limit. 
I will also present the status of $B_K$ and discuss its error budget, 
with a particular emphasis on the perturbative error.
Finally I will review the matrix elements of 
neutral kaon mixing beyond the standard model and 
will argue that the discrepancy observed by different 
collaborations could be explained by the renormalisation 
procedure of the relevant four-quark operators.
}

\FullConference{The 8th  International Workshop on Chiral Dynamics \\
		 29 June 2015 - 03 July 2015\\
		 Pisa, Italy}

\begin{document}

\section{Introduction}

Kaon physics has played and continues to play a central role 
in particle physics: 
CP violation, precision tests of the standard model, constraints 
on beyond-the-standard-model theories, etc.  
The recent progress achieved on the theoretical side, 
and in particular by the lattice community 
are improving drastically our theoretical understanding 
and open the door to new  phenomenological studies. 
I review here the 
lattice computations related to $K\to\pi\pi$ decays and neutral kaon mixing. 
For more general recent reviews on lattice flavour physics see for example~\cite{DeTar:2015orc, DellaMorte:2015rua}.

\section{$K\to\pi\pi$ decays and Lattice QCD}
Various nice reviews are available on the subject, see for example
~\cite{Lellouch:2011qw}. 
Let me just recollect some basic facts about $K\to\pi\pi$ phenomenology,
Assuming isospin symmetry, the decays $K\to\pi\pi$ can be written in terms of the amplitudes
\be
\label{eq:Ktopipi}
A\left [ K\to(\pi\pi)_I \right] = A_I {\rm e}^{i \delta_I}
\ee
where $I$ denotes the isospin of the two-pion state, either
$0$ or $2$, and $\delta_I$ is the corresponding strong phases.
The parameters of indirect (resp. direct) CP violation, $\varepsilon$
(resp. $\varepsilon'$)
are given by
\bea
\varepsilon &=& \frac{A\left[ K_L \to (\pi\pi)_0 \right]}{A\left[ K_S \to (\pi\pi)_0 \right]} 
\;.
\\
\frac{\varepsilon'}{\varepsilon}
&=&
{\frac{1}{\sqrt{2}}}
\left( 
\frac{A\left[ K_L \to (\pi\pi)_2 \right]}{A\left[ K_L \to (\pi\pi)_0 \right]}
-
\frac{A\left[ K_S \to (\pi\pi)_2 \right]}{A\left[ K_S \to (\pi\pi)_0 \right]}
\right)
\eea
The first measurement of $\epsilon$ is the well-known discovery of indirect CP violation 
due to Christenson, Cronin, Fitch and Turlay~\cite{Christenson:1964fg} in 1964,
for which Cronin and Fitch were awarded a Nobel prize.
$\epsilon'$ has a long experimental history as it took a tremendous effort
to measure direct CP violation. The final measurements are due to NA48 at Fermilab 
and KTeV at CERN ~\cite{Fanti:1999nm,AlaviHarati:1999xp}, the averages read
\bea
|\varepsilon| &=& 2.228(11) \times 10^{-3}\;,\\
{Re} \left(\frac{\varepsilon'}{\varepsilon}\right) &=&  16.6(2.3)\times10^{-4}  \;.
\eea

In a theoretical approach, 
the standard framework to study $K\to\pi\pi$ decay is the $\Delta S=1$
effective Hamiltonian obtained after integrating out 
the heavy degrees of freedom.
In the three-flavour theory, it reads
(see for example \cite{Buchalla:1995vs}, \cite{Cirigliano:2011ny})
\be
\label{eqH}
H_W = \frac{G_F}{\sqrt{2}}V_{us}^*V_{ud}\sum_{i=1}^{10} \bigl[z_i(\mu) + \tau y_i(\mu)\bigr] Q_i(\mu)\;,
\label{eq:H_W}
\ee
where $G_F$ is the Fermi constant.
The short-distance effects, which can be computed in perturbation theory are 
factorised into the so-called Wilson coefficients,
$y_i,z_i$ 
whose expression can be found in~\cite{Buchalla:1995vs}.
$V_{ij}$ are CKM matrix elements, $\tau = V_{ts}^* V_{td}/ V_{us}^* V_{ud}$
and $\mu$ is an energy scale which can be thought 
as a cut-off. 
The four-quark operators $Q_i$ can be obtained from Feynman diagrams
and after operator-product-expansion they can be expressed as
\vspace{0.5cm}\\
\begin{minipage}{3cm}
Current-Current: \\
\phantom{Current-Current}
\vspace{1cm}\\
QCD Penguins: \\
\vspace{3cm}\\
EW Penguins: \\
\vspace{3cm}\\
\end{minipage}
\begin{minipage}{12cm}
\bea
Q_1 &=& ({\bar s^i} \gamma_\mu(1-\gamma_5) d^i) ({\bar u^j} \gamma_\mu(1-\gamma_5) u^j) \vspace{1cm}\\
Q_2 &=& ({\bar s^i} \gamma_\mu(1-\gamma_5) d^j) ({\bar u^j} \gamma_\mu(1-\gamma_5) u^i) \\
%
\nonumber\\ 
Q_3 &=& ({\bar s^i} \gamma_\mu(1-\gamma_5) d^i) \sum_{q=u,d,s} ({\bar q^j} \gamma_\mu(1-\gamma_5) q^j) \\
Q_4 &=& ({\bar s^i} \gamma_\mu(1-\gamma_5) d^j) \sum_{q=u,d,s} ({\bar q^j} \gamma_\mu(1-\gamma_5) q^i) \\ 
Q_5 &=& ({\bar s^i} \gamma_\mu(1-\gamma_5) d^i) \sum_{q=u,d,s} ({\bar q^j} \gamma_\mu(1+\gamma_5) q^j) \\
Q_6 &=& ({\bar s^i} \gamma_\mu(1-\gamma_5) d^j) \sum_{q=u,d,s} ({\bar q^j} \gamma_\mu(1+\gamma_5) q^i) \\ 
\nonumber\\ 
Q_7   &=& ({\bar s^i} \gamma_\mu(1-\gamma_5) d^i) \sum_{q=u,d,s} e_q({\bar q^j} \gamma_\mu(1+\gamma_5) q^j) \\
Q_8   &=& ({\bar s^i} \gamma_\mu(1-\gamma_5) d^j) \sum_{q=u,d,s} e_q({\bar q^j} \gamma_\mu(1+\gamma_5) q^i) \\ 
Q_9   &=& ({\bar s^i} \gamma_\mu(1-\gamma_5) d^i) \sum_{q=u,d,s} e_q({\bar q^j} \gamma_\mu(1-\gamma_5) q^j) \\
Q_{10} &=& ({\bar s^i} \gamma_\mu(1-\gamma_5) d^j) \sum_{q=u,d,s} e_q({\bar q^j} \gamma_\mu(1-\gamma_5) q^i) 
\eea
\end{minipage}
\vspace{0.2cm}\\
The matrix element of these four-quark operators capture the 
strong dynamics of the theory. 
We have neglected the operators which emerge from the electric and magnetic dipole part
of the electromagnetic and QCD penguins. (See
the talk by V.Lubicz at Lattice'14 and~\cite{Lubicz:2014qfa}
for a recent lattice study by the ETM collaboration.)
These $10$ operators do not form a basis
of the $\Delta S=1$ four-quark operators in four dimensions,
as they are not linearly independent. 
Following~\cite{Blum:2001xb}, we build a $7$-operators basis:
\bea
Q_1' &=& 3 Q_1 + 2 Q_2 - Q_3 \\
Q_2' &=& \frac{1}{5} \left(  2Q_1 - 2Q_2 + Q_3 \right) \\
Q_3' &=& \frac{1}{5} \left( -3Q_1 + 3Q_2 + Q_3 \right) \\
Q_i' &=& Q_i\;, i\in\{5,6,7,8\} \;.
\eea
The $Q'_i$ falls into three different irreducible representations of $SU_L(3) \times SU_R(3)$:
$Q_1'$ transforms as a $(27,1)$, the QCD penguins\footnote{$Q'_{2,3}$ are actually 
combinations of current-current and QCD penguin operators. 
} $Q'_{2,3,5,6}$ as $(8,1)$  and 
the QED pengiuns as $(8,8)$.

Obtaining a reliable evaluation of the matrix elements $\la \pi\pi | Q_i' | K \ra $
is the most difficult part of the computation.
Since one needs a non-pertubative framework, lattice QCD 
is a natural candidate. 
In the last thirty years, many attempts have been
made to evaluate these matrix elements, using either effective
theories or lattice simulations (or combinations of both),
se for example 
\cite{Buras:2014maa,Cirigliano:2003nn, Bijnens:2009yr, Bijnens:2000im, Hambye:2003cy, Bardeen:1986vz, deRafael:1995zv} 
and reference therein.

From the lattice point of view, the first difficulty is to simulate the kinematic situation, 
in particular the final state made of two hadrons with non-vanishing momenta.
This problem was formalised in 1990 by Maiani and Testa who showed that the physical amplitudes could not be extracted 
from ``standard'' eulcidean lattice simulations~\cite{Maiani:1990ca}. 
An alternative  based on $\chi$PT was proposed in~\cite{Bernard:1985wf}:
the matrix elements of interests can be obtained from 
those of $K\to\pi$ and $K\to$ vacuum, which are numerically much simpler.
This indirect approach was first used for while, see for example~\cite{Gavela:1988ws,Noaki:2001un, Blum:2001xb}. 
However the conclusion of the extensive quenched studies~\cite{Noaki:2001un, Blum:2001xb}
is rather negative: extracting the matrix elements with a fully controlled error 
turned out to be very hard. One problem comes from the fact that $SU(3)$ 
$\chi$PT converges poorly at the kaon scale (see also \cite{Boucaud:2004aa})
~\footnote{
An interesting proposal based on Chiral-Scale Perturbation Theory 
has been presented at the conference by Lewis Tunstall,
~\cite{Crewther:2015dpa} see also \cite{Crewther:2013vea}. }.

What is now known as the Maiani-Testa no-go theorem was 
solved in an very elegant way by Lellouch and L\"uscher in~\cite{Lellouch:2000pv}.
The crucial point is that in finite volume the spectrum is discrete,
and the size of the box can be fine-tuned such that 
the pions will take the desired momentum.

\section{The $\Delta I =3/2$ channel}

We first consider the amplitude of $K\to(\pi\pi)_{I=2}$ decays, there are several simplifications in this channel,
most notably 
1. there is no disconnected diagram, and 2. only three operators contribute. 
The first realistic computation (with dynamical quarks, physical kinematics and nearly-physical pion mass) 
was performed by the RBC-UKQCD collaborations~\cite{Blum:2011ng,Blum:2012uk}
with Domain-Wall fermions, a discretisation of the QCD lagrangian
which preseves chiral-flavour symmetry almost exactly.

Although the method used in~\cite{Blum:2011ng,Blum:2012uk} is based on the Lellouch-L\"uscher approach,
an important ingredient is the Wigner-Eckart theorem, which tells us that 
the matrix elements of interest are related to those of the unphysical process $K^+ \to \pi^+ \pi^+$
(in the isospin limit).
Using a peculiar choice boundary conditions, these matrix elements 
(with physical momenta) can be extracted using standard lattice methods.
The first simulation was done at a single value of the lattice spacing 
($a^{-1}\sim 1.375$ GeV, ie $a\sim 0.1435$ fm) on the so-called IDSDR lattice
(ID) \cite{Blum:2014tka} with a pion mass of $140$ MeV.
(Strictly speaking this ``physical pion'' is partially quenched, the unitary pion mass was 
somewhat heavier: $170$ MeV).
In this work, the matrix elements are renormalised non-perturbatively , with 
the Rome-Southampton method~\cite{Martinelli:1994ty}. 
Since this lattice spacing is rather coarse, the renormalisation 
is first performed at a rather low value of the momentum scale ($\mu\sim 1.1 $ GeV).
In a second step, the same renormalisation factors are evaluated on finer lattices 
(called Iwasaki (IW) lattices) and the (universal) continuum scale-evolution matrix to $3$ GeV 
is obtained from ~\cite{Arthur:2011cn}, schematically:
\be
Z^{ID}(3 \,{\rm GeV}, a_{ID}) = 
\lim_{a_{IW} \to 0} \left[
  Z^{IW}(3 \,{\rm GeV},a_{IW})  \left( Z^{IW} (1.1 \,{\rm GeV} ,a_{IW}) \right)^{-1} 
\right]
Z^{ID}(1.1 \,{\rm GeV} , a_{ID}) \;.
\ee

More recently, the RBC-UKQCD collaborations have reported on 
$2+1$ lattice QCD simulations with physical pion masses~
\cite{Blum:2014tka}, 
which have been possible thanks to a new formulation of the Domain-Wall disctretisation
\cite{Brower:2012vk}.
These lattices have been used to improve on the determination of $A_2$:
the main source of error was the discretisation effects,
the new computation~\cite{Blum:2015ywa} 
involves two lattice spacings of $a\sim 0.011$ and $a\sim 0.084$ fm,
reducing the systematic error by roughly a factor $2$ for the real part and a factor $1.5$ for the imaginary part.
Thanks to these new lattice determinations, the current errors on the theoretical 
determination of $A_2$ are of the order of $10\%$.

\begin{figure}[t]
\begin{tabular}{cc}
\includegraphics[width=7cm]{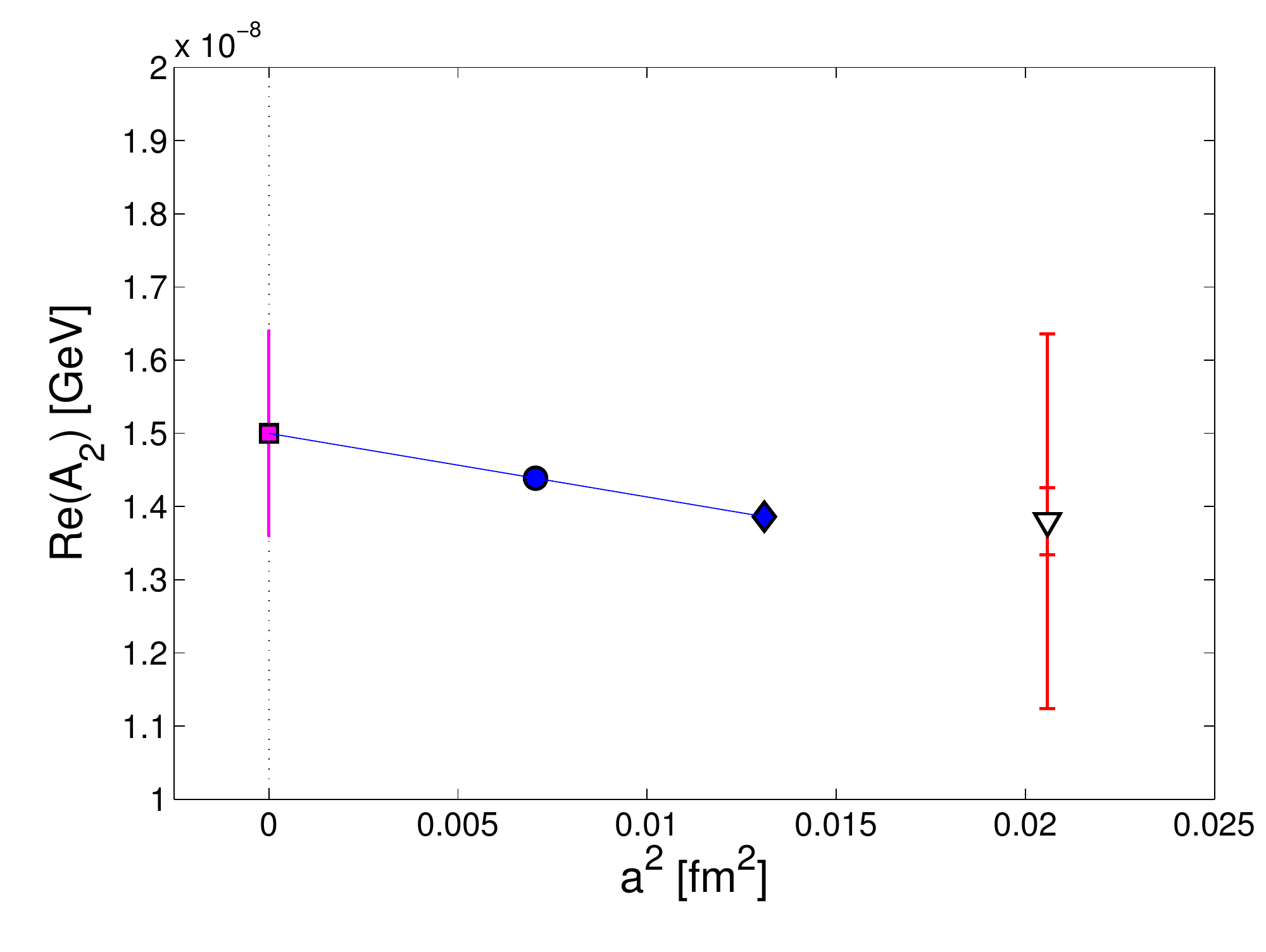} & 
\includegraphics[width=7cm]{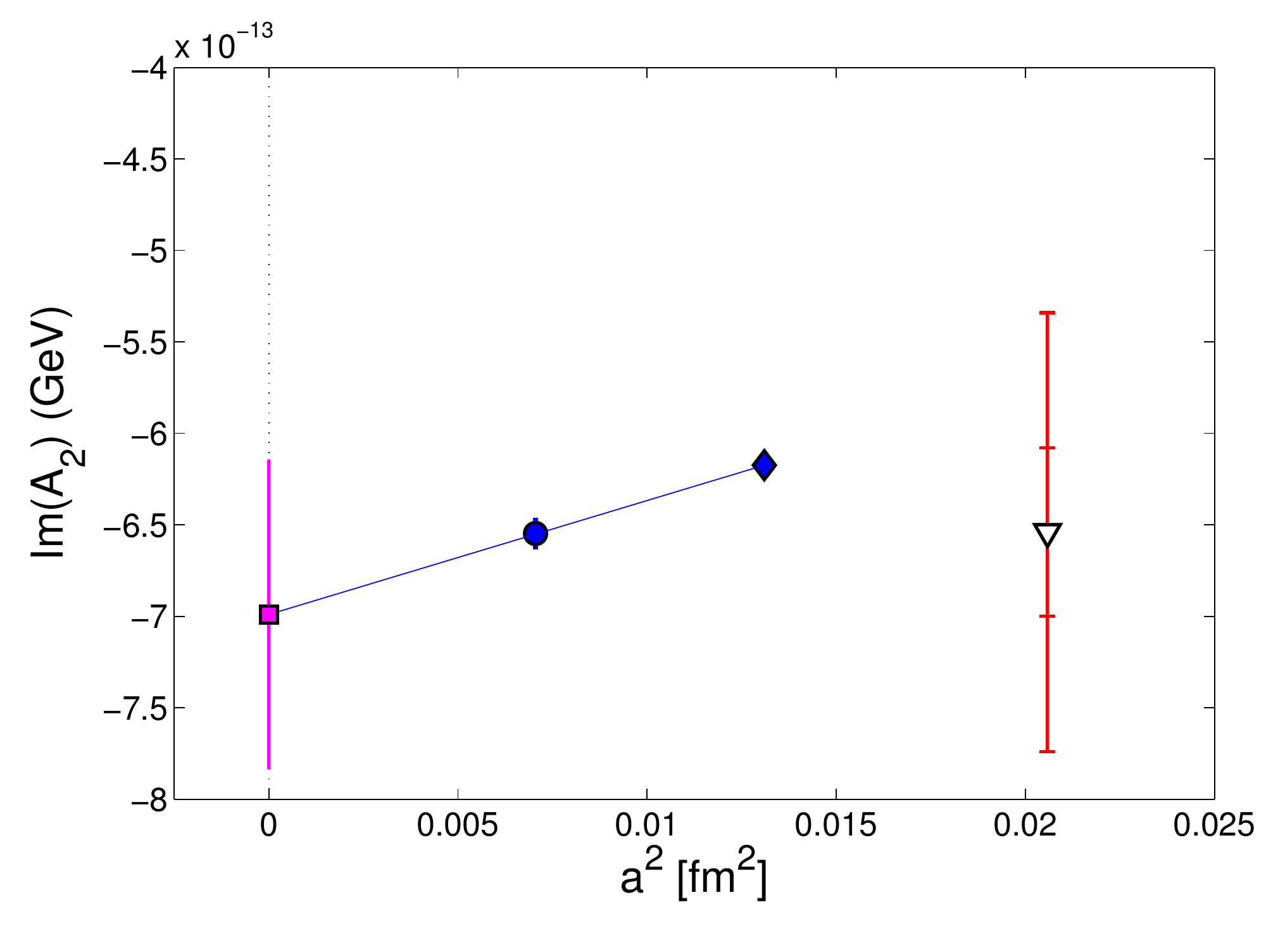}
\end{tabular}
\caption{Real and Imaginary part of $A_2=A\left[K\to(\pi\pi)\right]_{I=2}$. 
The triangle represents the 2012 computation on the IDSDR and the blue points
the 2014 determinations on the new ensembles (statistical error only), from which a continuum limit 
is extracted and shown in magenta (statistical and systematic errors combined). 
For the IDSDR points, 
we show both the statistical and the systematic error, largely dominated by 
the discretisation artefacts. 
}
\end{figure}

\section{Including the $\Delta I =1/2$ channel}

A complete determination of $A\left[K\to \pi\pi\right]_{I=0}$  has been a long-standing 
challenge for the lattice community. A first ``pilot'' computation with dynamical 
fermions was reported by RBC-UKQCD in 2011~\cite{Blum:2011pu}. 
This computation was unphysical in the sense that the amplitudes were computed
at threshold and the quark masses were heavier that the physical ones,
however all the required diagrams were determined (including the disconnected ones)
showing the numerical feasibility of the approach. The main remaining difficulty 
was to implement the physical kinematics, ie the ability of extracting the
matrix element of interests, with the pion states having the right momenta.
The Wigner-Eckart/boundary condition trick cannot used in the $\Delta =3/2$ 
channel cannot be used for the full computation as it violates isospin.
Instead, the RBC-UKQCD collaboration have generated new ensemble with $G-$ parity boundary conditions
\cite{Kim:2002np,Wiese:1991ku}, as reported by Christopher Kelly in a plenary 
session of Lattice 2015~\cite{CK2015},
see also the plenary review given by Andreas J\"uttner at the same conference~\cite{AJ2015}.
From a more technical point of view, this computation requires 
the evaluation of all-to-all propagators and noise reduction techniques.
The results read
\bea
Re(A_0) &=& 4.66(1.00)(1.21) \times 10^{-7} \rm GeV  \\ 
Im(A_0) &=& -1.90(1.23)(1.04) \times 10^{-11} \rm GeV \;
\eea
and the corresponding theoretical value for $\varepsilon'\varepsilon$ 
\be
 Re (\varepsilon'/\varepsilon)=1.38(5.15)(4.43)\times10^{-4}\;,
\ee
which is an approximate agreement($\sim2.1 \sigma$) with the experimental
value  $16.6(2.3)\times10^{-4}$. 
Rather than concluding that a significant deviation of the Standard Model
prediction has been found, we note that the error is much larger than experimental one.
From a phenomenological point of view, at this level of precision, 
these results do not invalidate  the standard model, neither do they rule the need 
for new-physics in $K\to\pi\pi$ decays.
The important point is that for the first time 
$\varepsilon'/\varepsilon$ has been computed with a full error budget,
all the different contributions of the seven linearly independent operators 
are computed with controlled errors and a precision which can be systematically improved. 
Now that the technology has been developed, 
reaching a precision of, say, $10\%$ should be possible in the close future.
In addition to reducing the statistical error, the simulation can be done
on finer lattices and extrapolated to the continuum limit.
Another systematic error is due to the truncation of the perturbation series 
(needed to compute the Wilson coefficients). 
The renormalisation was performed at a scale of $\mu\sim 1.5$ GeV
in order to keep the discretisation effects under control. 
Clearly this can be improved by running non-perturbatively to a higher scale, 
as done for the $\Delta I=3/2$ channel.  
Reducing the theoretical error on the matrix elements of $O_i$ (and therefore on n $\varepsilon'/\varepsilon$ )
will provide a crucial test of the standard model, 
indeed we might actually see signs of new physics. 
It is also worth noting that another computation (done at threshold) 
has been done 
with Wilson fermions~\cite{Ishizuka:2015oja}.

\section{The $\Delta I=1/2$ rule }
The ``$\Delta I=1/2$ rule'' refers to the fact that the $I=0$ channel is favoured over 
the $I=2$ channel by important factor $1/\omega$ defined by 
\be
\label{eq1}
\omega = \frac{A\left[ K_S \to (\pi\pi)_2 \right]}{A\left[ K_S \to (\pi\pi)_0 \right]} 
\ee
Experimentally this number is of order $\omega\sim 1/22$ whereas one would naively expects 
$1/2$~\cite{Gaillard:1974nj,Altarelli:1974exa}. 
The question whether or not the remaining factor of $\sim 10$ can be explained entirely by 
some surprisingly large QCD effects has been a very-long standing puzzle. It 
also shows the need for a better understanding of the non-perturbative regime.
Several attempts to study the $\Delta I=1/2$ rule on the lattice
have been made.
For example, an ongoing project based on the role of the charm quark has been developed 
in \cite{Endress:2014ppa,Hernandez:2008ft,Giusti:2006mh,Giusti:2004an},
see also ~\cite{Pena:2004gb}.\\

In 2013, the RBC-UKQCD collaborations reported on a study of the origin of this
enhancement ~\cite{Boyle:2012ys}. The amplitude $A_2$ was computed with physical kinematics
whereas $A_0$ was computed at threshold. 
\begin{figure}[t]
\begin{center}
\begin{tabular}{cc}
\includegraphics[width=5cm]{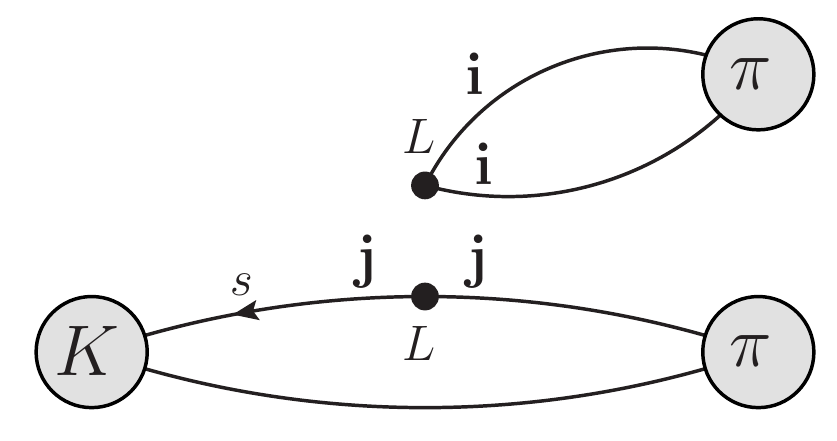} &
\includegraphics[width=5cm]{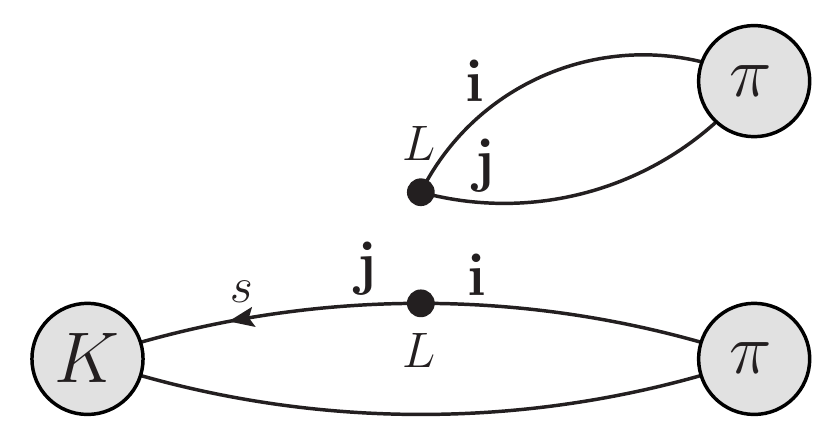}\\
\end{tabular}
\caption[]{
The dominating contribution to the real part of the 
amplitude $A_2$ of $K\to\pi\pi$ 
is proportional to sum of the two contractions 
$\circled{1}$  and $\circled{2}$.
The two contractions differ by their colour structure, 
as indicated by the colour indices $i$ and $j$.
The label $L$ stands for the left-handed
structure $\gamma_\mu(1-\gamma_5)$. 
}
\label{fig:contraction}
\end{center}
\end{figure}
The real part of $A_2$ is largely dominated by single four-quark operator
(the contributions of the electroweak penguins are negligible with 
respect to the tree-level diagram).
This operator has a ($V-A) \times (V-A) $ Dirac structure and transform as a $(27,1)$.
Two contractions $\circled{1}$ and  $\circled{2}$ contribute, they differ by their colour structure,
as shown in figure~\ref{fig:contraction}. 
The conventions are such that the real part of $A_2$ 
can be approximated by sum 
$\circled{1}+\circled{2}$
which are shown in figure~\ref{fig:contraction}.
The naive expectation is that $\circled{2}\sim \frac{1}{3} \circled{1}$.
However the observation made in~\cite{Boyle:2012ys} 
is that $\circled{2}\sim -0.7\circled{1}$. Therefore, there is important
cancellation in the numerator of Eq.(\ref{eq:omega}) which is completely 
unexpected from the naive factorisation framework.
Similarly the main contribution to $Re( A_0)$
is proportional to $2  \circled{1} -   \circled{2}$.
Hence, the aforementioned relative sign between  $\circled{1}$ and $\circled{2}$ 
also contributes to enhancement in the denominator of $\omega$ (compared to the naive 
expectation).
\be
\label{eq:omega}
\omega \sim \frac{ {Re}( A_2) }{ { Re} (A_0)} \sim 
\frac{\circled{1}+ \circled{2} }{2\circled{1}- \circled{2}}
\ee
The two recent lattice computations, the threshold one~\cite{Ishizuka:2015oja}
and the one with physical kinematics~\cite{Bai:2015nea}
also observe this sign difference, which seems to be at the origin of the $\Delta I=1/2$ rule.
However, in order to confirm that the $\Delta I=1/2$ effect
is a pure non-perturbative QCD effect, 
a little bit of patience is required as the precision on
$A_0$ has to be improved. 
The theoretical error affecting the amplitudes is expected to decrease
by a factor two in the next couple of years, 
it is very likely that we will then have the answer to this question.
Note that this sign also discussed in~\cite{Lellouch:2011qw,Pich:1995qp},
see also. ~\cite{Bardeen:1986vz,Buchalla:1995vs,Cirigliano:2011ny}.

\section{Neutral kaon mixing and indirect CP violation in the Standard Model} 

\begin{figure}[!h]
\begin{center}
\includegraphics[width=5cm]{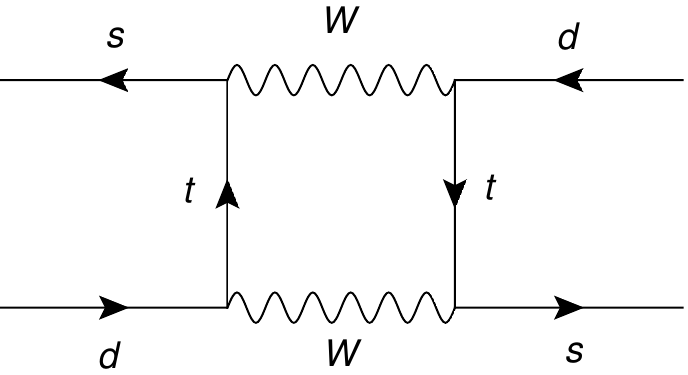}
\caption[]{Box diagram contributing to $K^{0}-\bar{K}^{0}$
mixing in the SM.} 
\label{fig:box}
\end{center}
\end{figure}

In the Standard Model picture, neutral kaon mixing 
is dominated by $W$-exchange box diagrams as illustrated in figure~\ref{fig:box}, 
a well-known loop-suppressed flavour changing neutral current.
By performing an operator product expansion, 
one can factorise the long-distance effects into  
the matrix element of a four quark operator:
\be
\label{eq:O1}
\la \Kb | O^{\Delta S=2}_1 | \K \ra
=
\la \Kb | 
(\overline s_i \gamma_\mu (1-\gamma_5) d_i)\,
(\overline s_j  \gamma_\mu (1-\gamma_5) d_j) \, 
 | \K \ra
\;.
\ee 
Clearly, because of the W-exchange, the Dirac structure is 
``$(\textrm{Vector-Axial})\times (\textrm{Vector-Axial})$ ''. 
It is quite remarkable that only one four-quark operator contribute:
even turning on QCD does not add any Dirac structure because 
the operator given in eq.~\ref{eq:O1} is invariant under 
Fierz re-arrangement.
In a (continuum) massless renormalisation scheme, 
it does not mix with other four-quark operators, nor with lower dimensional
operators. It is also the case on the lattice if chiral symmetry is preserved.

Once the considered matrix element has been computed non-perturbatively using lattice techniques, 
its result is combined 
with the value of the Wilson coefficient $C(\mu)$ of continuum perturbation theory and 
experimental observables, such as the mass difference $\Delta_{M_K}= m_{K_L}-m_{K_S}$ 
and 
$\epsilon_K$ to obtain important constraints on the CKM matrix elements. 
Schematically, one obtains
\be
\label{eq:epsilonSM}
\epsilon_K = C(\mu) \times \la \Kb | O^{\Delta S=2}_1 | \K \ra (\mu) 
\times {\cal F}(V_{ij}^{CKM},m_K,f_K,\Delta M_K,\ldots) \;,
\ee
where ${\cal F}$ is a known function of the CKM factors and of 
well-measured quantities.

A convenient parametrisation of this operator is the well-known bag parameter $B_K$,
\be
\label{eq:BK}
B_K(\mu) \equiv \frac{\la \Kb | O^{\Delta S=2}_1 | \K \ra (\mu)}{\frac{8}{3}f_K^2 m_K^2}.
\ee 
where $f_{K^-}=156.1$ MeV and $\mu$ is a renormalisation scale, 
usually $2$ or $3$ GeV. 
On the lattice, the bare bag parameter can be obtained from a ratio of correlators
\be
\label{eq:rbare}
r_{B_K}^{bare} =
\frac{ \la P^\dagger(t_2) O^{\Delta S=2}_1(t_o)  P^\dagger(t_1) \ra}
{   \la P^\dagger(t_2) A_0(t_o)\ra  \la A_0(t_o) P^\dagger(t_1) \ra }
\ee
where $P^\dagger$ create a light-strange pseudo-scalar particle 
(which would be a kaon in the continuum 
at the physical value of the quark masses) and $A_0$ is the time component
of the corresponding axial current. 
Up to some numerical factors, the bare bag parameter $B_K^{bare}$ 
is obtained from fitting $r_{B_K}^{bare}$ in the asymptotic 
euclidian time region $t_2 \ll t_o \ll t_1$

$B_K$ is a standard lattice quantity, nowadays it is computed with an accuracy 
of a few percents~\cite{Aoki:2010pe}\cite{Durr:2011ap}\cite{Constantinou:2010qv}.
The FLAG 2013 average for $N_f=2+1$ 
is 
\bea
\hat B_K &=& 0.7661(99) \;,\qquad N_f = 2+1 \;,
\eea
it is largely dominated by the BMWc result $\hat B_K = 0.773(8)_{stat}(3)_{syst}(8)_{PT}$.
The other references are 
\cite{Bae:2013lja, Arthur:2012opa, Laiho:2011np, Bae:2011ff, Durr:2011ap, Bae:2010ki, Aubin:2009jh, Antonio:2007pb,Allton:2008pn}.

Let us consider the dominant sources of error: 
FLAG 2013 explains that the total error of $99\sim 1.3\% $ can be roughly seen as the 
combination of $0.4\%$ statistical and $1.2\%$ systematic, mainly due to perturbation theory:
$(33)_{stat}+(93)_{syst}$. 
Although $B_K$ is extracted and (in most cases) 
renormalised non-perturbatively on the lattice,
perturbation theory is used to convert the result to the renormalisation-group-invariant
(RGI) quantity $\hat B_K$, or alternatively to $\msbar$. 
Naturally, different collaborations estimate the  perturbative error in different ways,
and this estimation is of course affected by some subjective judgement. 
Indeed this error changes by a factor two or three depending 
on the estimation.
In its 2013 review, it seems that FLAG chose an uncertainty 
very close to the one quoted by BMW ($1\%$), whereas RBC-UKQCD
quoted an error of $\sim 2\%$, based on a multiple-scheme evaluation.
Actually by changing the intermediate schemes,  RBC-UKQCD find 
that the results change by $8\%$ if the matching is done at $\mu=3\GeV$
and by $12\%$ if $\mu=2\GeV$.
The current situation is illustrated 
on table~\ref{table:BK}, where we have collected 
the most recent computation of $B_K$.
The importance of this perturbative error 
can be made clear by looking at, for example, 
the result obtained by
RBC-UKQCD~\cite{Blum:2014tka}
\bea
\hat B_K  &=& 0.7499 (24)_{stat}(150)_{PT} \\
B_K^{(\s{q},\s{q})}(3 {\rm GeV}) &=& 0.5341(18)_{stat} 
\eea
where the first error is statistical (however it is much 
larger than the other errors on $B_K^{bare}$) and the second error
is the systematic error on the renormalisation, largely
dominated by the perturbative matching.
This contrasts with $B_K^{(\s{q},\s{q})}$ which is fully non pertutbative,
renormalised in the SMOM $(\s{q},\s{q})$-scheme at the scale $\mu=3$ GeV.
Without this perturbative error, the error would be of around $0.3\%$.

In the future, the lattice community will probably have to find an agreement
on how to estimate this uncertainty as it is the dominant one. 
Another way to improve the situation
is obviously to reduce this perturbative error. 
One way would be to compute explicitly the next order in perturbation theory.
The matching coefficient is currently known at next-to-leading order.
Going further requires to determine the matching coefficient 
at the two-loop level (three-loop  anomalous dimension).
Alternatively, on could perform the matching at a higher scale;
this could be achieved by computing the running non-perturbatively, for example 
using the Schr\"odinger functional~\cite{Papinutto:2014xna}
or a (S)MOM-scheme, as presented in ~\cite{Frison:2014esa}.

\begin{table}[t]
\begin{center}
\begin{tabular}{cccl}
Collaboration & $N_f$ & Discretisation & Result \\
RBC-UKQCD~\cite{Blum:2014tka} & $2+1$ & Domain-Wall &  $0.5293(17)_{stat}(106)_{PT}$ \\
SWME~\cite{Jang:2015sla} & $2+1$   & Staggered & $0.518(3)_{stat}(26)_{syst}$  \\
ETM\cite{Carrasco:2015pra}        & $2+1+1$ & Twisted Mass &  $0.506(17)_{stat+syst}(3)_{PT}$ \\
\end{tabular}
\end{center}
\caption{Collection of recent results for $B_K^{\msbar}(3{\rm GeV})$. 
In~\cite{Blum:2014tka}, the first error is statistical but is 
much larger than the systematic ones, except for the perturbative error
which is written as $(PT)$. For the latter, it is interesting to note that
RBC-UKQCD quotes an error of $2\%$ whereas ETM quotes $0.6\%$.
In~\cite{Jang:2015sla} this error is not distinguished from other
uncertainties due to the renormalisation procedure.
BMWc result is not quoted here as it uses different convention, 
but quotes a perturbative error of $1\%$. 
This perturbative error is discussed in detail in the text.
}
\label{table:BK}
\end{table}

\section{Neutral kaon mixing Beyond the Standard Model}

In the standard model operator, only the operator $(V-A)\times(V-A)$ contributes
to kaon oscillations, because the process can only occur through $W$ exchanges. 
Beyond the standard model, we have to include new Dirac-colour structures,
as for example both left-handed and right-handed currents can contribute to 
$K^0$-$\bar{K}^0$ mixing (and therefore to 
$\epsilon_K$).  
Hence, in addition to $O_1$ introduced in eq.\ref{eq:O1},
one introduces new $\Delta S=2$ four-quark operators.
A possibility (the so-called SUSY-basis) is ~\cite{Allton:1998sm,Ciuchini:1998ix}
~\footnote{
An alternative basis is the one given by~\cite{Buras:2000if}
.}
\bea
O_2^{\Delta S=2}&=\;(\overline s_i (1-\gamma_5) d_i)\,(\overline s_j  (1-\gamma_5) d_j),\\
O_3^{\Delta S=2}&=\;(\overline s_i  (1-\gamma_5) d_j)\,(\overline s_j   (1-\gamma_5) d_i),\\
O_4^{\Delta S=2}&=\;(\overline s_i  (1-\gamma_5) d_i)\,(\overline s_j   (1+\gamma_5) d_j),\\
O_5^{\Delta S=2}&=\;(\overline s_i  (1-\gamma_5) d_j)\,(\overline s_j   (1+\gamma_5) d_i).
\eea

The BSM operators  $O_{i\ge2}^{\Delta S=2}$ mix pair-wise: $O_2^{\Delta S=2}$ with $O_3^{\Delta S=2}$ 
(in the chiral limit they transform under a $(6,\bar 6)$ irreducible 
representation of $SU(3)_L \times SU(3)_R$)
and  $O_4^{\Delta S=2}$ with $O_5^{\Delta S=2}$  (the corresponding group irrep being $(8,8)$). 
These four-quark operators appear in the generic effective $\Delta S=2$ Hamiltonian
\be
\label{eq:H}
H^{\Delta S=2} = \sum_{i=1}^{5} \, C_i(\mu) \, O_i^{\Delta S=2}(\mu) 
 + 
\,,  \sum_{i=1}^{3} \tilde C_i(\mu)
\ee
where the Wilson coefficient $C_i(\mu), \tilde C_i(\mu)$ 
depend on the details of the new-physics model under consideration
but the matrix elements $\la \Kb | O^{\Delta S=2}_i | \K \ra$ are model independent.
The operators $\tilde O_{1,2,3}$ are obtained from 
$O_{1,2,3}$ by replacing $(1-\gamma_5)$ by $(1+\gamma_5)$. 
In QCD with parity conserved, these operators are redundant and therefore discarded 
in the following.

{\em A priori}, one would expect that the relevant matrix elements
$\la \Kb | O^{\Delta S=2}_1 | \K \ra$ can be obtained with an accuracy
comparable to the one of the Standard Model one. 
However only few studies of the full set of BSM operators 
have been published and the history is quite interesting.
First of all, in the quenched approximation, the results
from~\cite{Babich:2006bh} obtained with Ginsparg-Wilson fermion (which exhibit an exact
chiral-flavour symmetry) and non-perturbative renormalisation 
were very different from the previous study, done with 
tree-level $O(a)$-improved Wilson fermions~\cite{Donini:1999nn}. 
The difference was attributed to the renormalisation.

The first computation performed with dynamical fermions was reported by RBC-UKQCD~\cite{Boyle:2012qb} 
in 2012 and was done with $N_f=2+1$ Domain-Wall fermions.  
It was followed shortly by a $N_f=2$ twisted-mass computation of the ETM collaboration, 
done with several lattice spacings~\cite{Bertone:2012cu} and these two first computations 
are in reasonable agreement (slightly more than$2\%$ in the worse case)
In 2013, the SWME collaboration reported 
their results, obtained with $N_f=2+1$ flavours of improved staggered 
fermions~\cite{Bae:2013tca}. 
A noticeable disagreement 
with the previous studies was found for two of the matrix elements
($O_4$ and $O_5$ of the SUSY basis).
Very recently, the ETM collaboration have repeated their 
computation with $N_f=2+1+1$ flavours (using again twisted mass QCD),
and essentially confirmed their $N_f=2$ results~\cite{Carrasco:2015pra}
(although for $B_5$ the agreement is only within $\sim3\sigma$).
Even more recently, SWME have extended their study by adding more ensembles,
\cite{Jang:2015sla}
improving the extrapolation to the physical point,
and they confirmed the disagreement with the other collaboration.
Since the results have been extrapolated to the continuum limit,
one does not expect  that  the discretisation used (Domain-Wall, Twisted-Mass, 
or Staggered) is responsible of the discrepancy. 
A first suspect is the renormalisation procedure. ETM and RBC-UKQCD employ 
a non-perturbative method, based on the Rome-Southampton method\cite{Martinelli:1994ty},
whereas SWME uses a perturbative matching. 
Another possibility could an underestimated systematic error due to a chiral log.
Although this sounds more unlikely it could be investigated,
in a first step by extrapolating quantities which 
have different chiral behaviour (the $B's$, the $R's$ and the $G's$)
and comparing the results. Eventually these quantities will be
computed directly at the physical value of the pion mass, as done for the 
Standard Model contribution.

\clearpage\newpage
{\bf Normalisation.}
The matrix elements of these four-quark operators are usually 
given in terms of the so-called Bag-parameters, 
\be
B_i(\mu) =\frac{ \la \Kb |O_i(\mu)| \K \ra } { \la \Kb |O_i(\mu)| \K \ra}_{VS}  \;,
\ee
where $VS$ is the vacuum-saturation approximation.
On the lattice, this is by achieved by computing a ratio of three-point function
over a product of two-point functions, such as $r^{bare}_{B_K}$ 
defined in eq.~(\ref{eq:rbare}). 
In the case of the Standard Model operator, the denominator is known 
in terms of physical quantities $f_K$ and $m_K$, as shown in eq.~(\ref{eq:BK}).
This normalisation is convenient because the bags parameters are dimensionless, 
the numerator and the denominator are very similar, therefore systematic 
errors are likely to cancel out in the ratio, and because the denominator is 
known in terms of physical quantities. 
However for the BSM operators, the corresponding vaccum saturation involve
matrix element of the pseudo-scalar density, and on the lattice 
the corresponding $r^{bare}_{B_{i>1}}$ usually contains the product of 
pseudo-scalar pseudo-scalar two-point function 
$\la P^\dagger(t_2) P(t_o)\ra \la P(t_o) P^\dagger(t_1) \ra $, 
which is approximated by $m_K^4f_K^2/(m_s+m_d)^2$.
First, this is only an approximation, but also the fact 
that quark masses appear 
imply new ambiguities such as scale and scheme dependencies. 
Numerically, we also find the ratio of three-point functions
is better determined than the ratio three-point over two point functions.

This problem is well known and several alternatives were proposed in the literature, 
see for example~\cite{Donini:1999nn}.
Let us mention explicitly 
the solution proposed in
~\cite{Babich:2006bh}. 
Denoting by $P$ the simulated pseudo-scalar particle (kaon) 
of mass $m_P$ and decay constant $f_P$, the $R$'s are defined by
\bea
\label{eq:R}
{\cal R}_i\left (\frac{m_P^2}{f_P^2}, \mu, a^2 \right)= 
\left[ {\frac{f_K^2}{m_K^2}} \right]_{\rm expt}
\left[ {\frac{m_P^2}{f_P^2} } { \frac{\langle \Pb| O_i(\mu) | \P \rangle }{\langle \Pb|O_1(\mu) | \P \rangle } }\right]_{\rm latt} \;. 
\eea
such that at the physical point $m_P=m_K=m_K^{\rm expt}$ and $a= 0$
\bea
\label{eq:Rphys}
R_i (\mu) = 
{\cal R}_i 
\left( \frac{m_K^2}{f_K^2}, \mu, 0 \right) &=& 
{
\frac{\langle \Kb| O_i(\mu) | \K \rangle}
{\langle \Kb|O_1(\mu) | \K \rangle} }
\eea
is directly proportional to the ratio of a BSM contribution to the SM one.\\

Another possibility, advocated for example in~\cite{Bae:2013tca}, 
is to define products and ratios of bag parameters (called $G$) 
such that the leading  chiral logarithms cancel out.
This cancellation occurs at best at every order of the chiral expansion, 
or in the worst case to next-to-next-to-leading order (NNLO). Such quantities were introduced
in \cite{Becirevic:2004qd} for $SU(3)$ chiralperturbation theory and expanded later in the context of
$SU(2)$ staggered chiral perturbation theory in \cite{Bailey:2012wb}.
The problem of the normalisation ambiguity is absent for the ratios, 
but still there for the product. 
However the advantage is that the chiral extrapolations are hugely simplified.
Having different normalisations and fit Ans\"astze help to a better control of the systematic 
error. One can for example extract the $B$ directly with some chiral logs, 
of reconstruct them from the linearly extrapolated $G$. 
\vspace{0.5cm}

\begin{table}[tb]
  \begin{center}
$$
    \begin{array}{c| c |c } 
\hline
      R^{\rm SUSY}(3\,\GeV)  & {\rm SMOM}-\gamma_\mu   & \msbar  \\
\hline
R_2 & -19.11(43)(19)(25)           &  -19.48(44)(20)(25)      \\
R_3 &  \ph{-1} 5.76(14)(15)(07)    & \ph{-1} 6.08(15)(16)(08) \\
R_4 &  \ph{-1}40.1(08)(17)(09)     & \ph{-1}43.1(09)(18)(10)  \\
R_5 &  \ph{-}11.13(21)(79)(25)     & \ph{-}10.99(20)(78)(25)  \\
\hline
    \end{array}
$$
    \caption{RBC-UKQCD preliminary results for the BSM $\Delta S=2$ ratio $R$ in the SUSY basis,
in the $\gamma_\mu$-SMOM scheme and in $\msbar$  at $3\,\GeV$. 
The quantities $R$ are the ratios of the BSM matrix elements over the SM contributions.
Errors are statistics, discretisation, chiral, respectively. 
}
\label{table:R}
  \end{center}
\end{table}

{\bf New RBC-UKQCD results (preliminary) and tentative explanation of the disagreement.}

I am presenting now a work in progress with the RBC-UKQCD collaborations,
preliminary results have been presented by R.J.Hudspith at Lattice 2015~\cite{RJH2015}. \vspace{0.1cm}\\

{\bf Possible explanation.} 
In order to understand the source of the disagreement, 
we have extended our previous study~\cite{Boyle:2012qb} 
in several ways, most notably
1.  by adding another lattice spacing 2. by investigating 
new non-perturbative renormalisation (NPR) SMOM schemes, 
in the spirit of the schemes introduced in~\cite{Sturm:2009kb}.
Our main results are presented in Table~\ref{table:R}, in a SMOM scheme,
and in $\msbar$ (we would like to thank Christoph Lehner for computing
the conversion factors for $B_2$ and $B_3$). 
A comparison of results for the bag parameters can be found in Table~\ref{table:B}.
Although our error budget is not complete yet, we find that if we use 
the standard RI-MOM scheme proposed in~\cite{Martinelli:1994ty} 
and match to the $\msbar$ scheme defined in 
\cite{Buras:2000if}, 
our results are in a decent agreement with ETMc. 
Surprisingly enough. if we use a SMOM scheme, our results are much closer 
to the results quoted by SWME, for which the renormalisation is performed perturbatively. 
The SMOM schemes are known to be superior to standard RI-MOM schemes: they behave better non-perturbatively 
in the infrared (the pion pole contamination is suppressed because of the 
absence of exceptional channel) and perturbatively. 
Our suspicion is that the procedure employed to remove the pion pole contamination
(needed in the RI-MOM case but absent for the SMOM schemes) could also affect the 
ultraviolet behaviour, see for example \cite{Lytle:2014tsa}. 
The systematic errors associated with this procedure 
are very hard to estimate, they could have been underestimated.
We do not advocate to use the bag parameters for our central values, we only show 
them in order to compare with collaborations. 
We find that the $R$'s have smaller systematic and give much more reliable results.
On the negative side, we find that - regardless of the normalisation - 
the discretisation effects are larger than expected. 
Although these artefact are moderate (we quote $7\%$ from the $a^2$ slope in the worse case)
they are larger than what is usually found with Domain-Wall fermions.
In the future, we are planning to add a third lattice spacing in order
in order to have a better handle on these lattice artefacts.

\begin{table}[tb]
  \begin{center}
$$
\small
    \begin{array}{c| c | c | c | c  |c c} 
    &   \rm{ETM'12} & \quad  \rm{ETM'15} \quad & \rm{RBC-UKQCD'12} & \quad \rm{SWME'15} \quad& \multicolumn{2}{c}{\qquad {\rm RBC-UKQCD'15 (prelim.)}}  
\\
interm. & & & & & & \\
scheme                 &     RI-MOM      &    RI-MOM              & RI-MOM         & 1-loop &    RI-SMOM               & RI-MOM \\
      \hline 
      B_2         &  0.47(2)   & 0.46(3) &    0.43(5)            &  0.525(1)(23)  \;     & 0.488(7)(17)(2)  & 0.417(6)(2)(2)   \\ 
      B_3         &  0.78(4)  & 0.79(5) &     0.75(9)            &  0.772(5)(35)  \;     & 0.743(14)(64)(3) & 0.655(12)(44)(2) \\ 
      B_4         &  0.75(3)  & 0.78(5) &     0.69(7)            &  0.981(3)(61)  \;     & 0.920(12)(12)(4) & 0.745(9)(28)(3)  \\ 
      B_5         &  0.60(3)  & 0.49(4) &     0.47(6)            &  0.751(8)(68)  \;     & 0.707(8)(34)(3)  & 0.555(6)(53)(2)  \\ 
    \end{array}
$$ 
    \caption{Comparison of the $B's$ at $3\,\GeV$ in the SUSY basis in the $\msbar$ scheme of \cite{Buras:2000if}.
      {\em We do not recommend to use the bag as a parametrisation of the BSM four-quark operators, we 
        only quote them in order to compare the results obtained by different collaborations.}
            In $\rm{ETM'12}$ and  $\rm{RBC-UKQCD'12}$ the renormalisation was performed in the $\rimom$ scheme 
      with exceptional kinematics whereas in $\rm{SWME'14}$ it is performed perturbatively.
      The new preliminary RBC-UKQCD results are computed using different intermediate schemes
      and clearly show that the intermediate scheme difference is much larger
      than expected (an $\alpha_s^2$ effect). 
      We argue that the renormalisation procedure is the cause of the disagreement observed 
      for $B_4$ and $B_5$ between the different collaborations and that it is due to some underestimated 
      systematic errors present in the $RI-MOM$ scheme. See the text for more detailed explanation. 
      The errors have been already combined, except for SWME'15 where
      the first errors are statistical and the second systematic.
      For the RBC-UKQCD'15, 
      the errors are statistics, discretisation and  chiral respectively. All the other errors 
    are already combined.}
  \end{center}
\label{table:B}
\end{table}

\section{Conclusions and outlook}
Kaon physics is experiencing a new youth (see for example \cite{Buras:2015hna});
to a great extent this is due to the impressive progress achieved recently by the lattice community.
The computation of $K\to(\pi\pi)_{I=2}$ is reaching a mature stage
and a first computation $K\to(\pi\pi)_{I=0}$ with physical kinematic
and complete error budget has recently been 
reported by the RBC-UKQCD collaboration.
The results of these computations have 
a important role to play in particle physics phenomenology . 
The $\Delta I=1/2$ puzzle seems to be explained by the non-perturbative effects
\cite{Boyle:2012ys}.
Regarding indirect CP violation, $B_K$ is now known with an impressive precision. 
The various investigations of the $\Delta S=2$ BSM operators
are converging, the discrepancies observed 
by several collaborations are likely to be due to systematic errors
affecting the non-perturbative renormalisation procedure 
in RI-MOM schemes. Although a careful study is required, 
the solution could be provided by the SMOM schemes, which have a much 
better behaviour. Future improvements will also require to match 
the lattice computation to phenomenology at a much higher scale
in order to decrease the error due to perturbation theory. 
I have presented here the new determinations of $K\to\pi\pi$ decay amplitudes
and neutral kaon mixing matrix elements, 
but there are other new interesting developments 
that I have not mentioned here,
such as rare kaon decys and the $K_L-K_s$ mass differrence
(see \cite{Christ:2015aha} and \cite{Christ:2015pwa} ). 
\\

\vspace{1.cm}
{\bf Acknowledgements-} 
I would like to thank the organisers for such an enjoyable conference
and for promoting my talk. In particular, 
the conveners of the Goldstone boson working group,
Mario Antonelli, Sebastien Descotes-Genon, Andreas J\"uttner, Emilie Passemar 
and Michele Viviani,
not only for the organisation, but also the physics discussion we had.
I am indebted to Lewis Tunstall for discussing his work on Kaon decay 
and Chiral perturbation theory. 
I also would like to thank my colleagues of the RBC-UKQCD collaboration 
and my lattice colleagues who shared their work with me, sometimes prior to publication.

\bibliography{biblio}{}
\bibliographystyle{h-elsevier}

\end{document}